\documentclass[twocolumn,10pt,titlepage]{article}

\usepackage{amssymb}
\usepackage{graphicx}
\usepackage{dcolumn}
\usepackage{bm}
\usepackage[sort&compress]{natbib}

\begin{document}

\title{Equation of state and elastic properties of face-centered-cubic FeMg alloy at ultrahigh pressures from first-principles}

\author{C. Asker, U. Karg{\'e}n, L. Dubrovinsky and I.A. Abrikosov}

\maketitle

\begin{abstract}
We have calculated the equation of state and elastic properties of face-centered cubic Fe and Fe-rich FeMg alloy at ultrahigh pressures from first principles using the Exact Muffin-Tin Orbitals method.
The results show that adding Mg into Fe influences strongly the equation of state, and cause a large degree of softening of the elastic constants, even at concentrations as small as 1-2 at. \%.
Moreover, the elastic anisotropy increases, and the effect is higher at higher pressures.
\end{abstract}


\section{Introduction} \label{Introduction}
The properties of iron at ultrahigh pressure have been studied extensively, both experimentally and theoretically. \citep{science316_1880,science295_313,nat422_58,nat424_1032,phrL95_245502,phrL99_165505, EPSL268_444, EPSL273_379,nat424_536,epsl254_227,science319_797,nat413_57}
Apart from a better understanding of one of our most important metals, one aim is to find the composition and structure of the Earth's core, 
which is likely to be made up of iron mixed with lighter elements.

For a light element to be considered as a possible component in the Earth's core, it should naturally alloy with Fe and be abundant within the Earth.
While it is well-known that Mg is abundant\citep{anderson}, until recently, magnesium was not considered as a possible candidate to be one of the light elements in the core, mainly because iron and magnesium differ too much in size to form alloys. 
However, it has been shown both theoretically and experimentally that at high pressures, the size mismatch of iron and magnesium has decreased enough so that FeMg alloys can form \citep{phrL95_245502,GeoPhysResLett32_L06313}.
At oxygen fugacity corresponding to Mg-MgO buffer, up to 10 $at. \%$ of Mg could be dissolved in hcp Fe at about 100 GPa pressure \citep{phrL95_245502}.
In presence of magnesiow{\"u}stite (Mg,Fe)O and dense silicate perovskite $(Mg,Fe)SiO_{3}$, 
about 1 $at. \%$ Mg was observed in hcp Fe \citep{GeoPhysResLett32_L06313}.

We acknowledge that the relatively high oxygen fugacity in the Earth interior, 
makes Mg a somewhat unlikely candidate for the main light element in the core \citep{phrL95_245502}. 
Still its presence has not been ruled out, and the study of its effect on the properties of Fe is motivated.

The elastic properties of iron rich FeMg alloys were recently reported for the hexagonal close-packed (hcp) \citep{EPSL271_221} and for body-centered cubic (bcc) \citep{PNAS106_15560} structures.
However, Mikhaylushkin \emph{et al.} have shown that the fcc phase of Fe can not be ruled out as a possible structure stable at conditions corresponding to the Earth's core. \citep{phrL99_165505}
Since the energy difference between the competing phases of $Fe$ at Earth core conditions is extremely small \citep{phrL99_165505}, we therefore believe it is important to investigate elastic properties for the face-centered cubic structure (fcc) of $Fe$ rich FeMg alloy. 
In particular, in a recent study \citep{FeNisubmitted} we have demonstrated that the elastic anisotropy of fcc $Fe$, estimated according to Hill's definition~ \citep{grimvall}, is much higher than for hcp $Fe$, and the effect persists in $FeNi$ alloys. \\

We report in this letter first-principles calculations of the equation of state and elastic properties of fcc $Fe$, $Fe_{98}Mg_{02}$, $Fe_{95}Mg_{05}$ and $Fe_{90}Mg_{10}$ at ultrahigh pressures.
The calculations have been carried out using the EMTO-CPA method \citep{phrL87_156401} (described below), which is a suitable tool for calculating elastic properties in alloys.

The paper is organized as follows: in Sec.~\ref{Theory} we outline the theory used for calculating elastic properties from first-principles by the EMTO-CPA method together with details of our calculations and in Sec.~\ref{Results} we present the equation of state, the single crystal and polycrystalline elastic properties  of fcc FeMg.


\section{Theory} \label{Theory}
In this work, we first calculated the total energy for a set of atomic volumes between $41.6$, and $78.0$ $Bohr^3$.
The results from these calculations were then used to find the equation of state (EOS) through the Birch-Murnaghan scheme~ \citep{phr71_809} (3rd order).
The EOS provides a relation between volume and pressure as well as allows us to study the pressure dependence of the bulk modulus. \\

In cubic structures, there are three independent elastic constants: $c_{11}$, $c_{12}$ and $c_{44}$.~ \citep{grimvall}
The $c^{\prime}$ elastic constant (also known as Zener's elastic constant) can be obtained from the orthorhombic structure with the following deformation~ \citep{levente_book}:
\begin{equation} \label{eq:fcc-c_prime-dist}
\mathbf{I} + \mathbf{D_{O}} = 
\left( \begin{array}{ccc}
1 + \delta & 0 & 0 \\
0 & 1 - \delta & 0 \\
0 & 0 & \frac{1}{1-\delta^{2}}
\end{array} \right)
\end{equation}
The change in total energy upon distortion is
\begin{equation} \label{eq:deltaE-fcc-c_prime}
\Delta E  = 2 V \cdot c^{\prime} \cdot \delta^{2} + \mathcal{O}(\delta^{4})
\end{equation}

Now the $c_{11}$ and $c_{12}$ can be calculated from the relations between $c^{\prime}$ and the bulk modulus:
\begin{equation} \label{eq:Bmod-cubic}
B = \frac{c_{11} + 2c_{12}}{3} 
\end{equation}
and
\begin{equation} \label{eq:c_prime-def}
c^{\prime} = \frac{c_{11} - c_{12}}{2}
\end{equation}
From Eqs.~(\ref{eq:deltaE-fcc-c_prime},\ref{eq:Bmod-cubic},\ref{eq:c_prime-def}) the $c_{11}$ and $c_{12}$ elastic constants can be obtained.

Next, the $c_{44}$ elastic constant can be obtained by performing monoclinic distortion~ \citep{levente_book}:
\begin{equation} \label{eq:fcc-c44-dist}
\mathbf{I} + \mathbf{D_{M}} = 
\left( \begin{array}{ccc}
1 & \delta  & 0 \\
\delta & 1 & 0 \\
0 & 0 & \frac{1}{1-\delta^{2}}
\end{array} \right)
\end{equation}
In this case the change in energy is:
\begin{equation} \label{eq:deltaE-fcc-c44}
\Delta E  = 2 V \cdot c_{44} \cdot \delta^{2} + \mathcal{O}(\delta^{4})
\end{equation}

In the calculations of both $c^{\prime}$ and $c_{44}$ the distortions used were $\delta = [ 0.00, 0.01, \ldots, 0.05 ]$. 
The elastic constants were then obtained from (linear) interpolations of the total energy as function of the \emph{square} of the distortion, $\Delta E_{tot} = \Delta E_{tot}(\delta^{2})$. 
The motivation for this procedure is that it simplifies the choice of suitable distortions, and is numerically preferable over interpolation of distortion.
In this work we only do volume-conserving distortions.
The reason for this is that we seek to calculate the elastic properties as functions of pressure, but in the computational setup we must use volume (rather than pressure) as input parameter. 
The EOS can then be used to obtain the elastic constants as functions of pressure.
Also, the total energy depends more strongly on volume than on the distortions~ \citep{phrB60_791}, indicating that it should be more stable numerically to do volume-conserving distortions whenever possible.\\

The elastic constants described above (except $B$) are single-crystal elastic constants. 
When doing measurements, one is often dealing with polycrystalline samples of a material, and hence the single-crystal constants should be averaged in a suitable way to facilitate comparison between experiments and theoretical calculations.
The polycrystalline shear moduli defined by Reuss and Voigt are \citep{grimvall}:
\begin{eqnarray}\label{eq:Gv-fcc}
G_{R}&=& \frac{5(c_{11}-c_{12}) c_{44}}{4 c_{44}+3(c_{11}-c_{12})} = \frac{5c^{\prime}c_{44}}{2c_{44}+3c^{\prime}} \label{eq:Gr-fcc}\\
G_{V} &=& \frac{c_{11}-c_{12}+3c_{44}}{5} = \frac{2c^{\prime} + 3c_{44}}{5} \quad ,
\end{eqnarray}
respectively.
In the case of cubic systems, the Reuss and Voigt bulk moduli are equal, and defined in Eq.(\ref{eq:Bmod-cubic}).

Further, the elastic anisotropy according to Hill is the weighted difference between the Voigt and Reuss shear moduli:
\begin{equation} \label{eq:anisotropy}
A_{VRH}=\frac{G_{V}-G_{R}}{G_{V}+G_{R}} \quad ,
\end{equation}
and hence is called the \emph{Voigt-Reuss-Hill} (VRH) anisotropy \citep{grimvall}. \\

All the calculations in this work have been carried out within Density-Functional Theory~ \citep{phr1363b,phr1404a} (DFT) using the \emph{Exact Muffin-tin Orbital Method} (EMTO)~ \citep{andersen-notes,CMS18_24,phrB64_014107}. 
The problem of disorder was treated with the Coherent Potential Approximation (CPA)~ \citep{phr156,phr175_747}, implemented in the EMTO-CPA \citep{phrL87_156401} method.
For exchange and correlation the Local-Density approximation (LDA), parametrized by Perdew, Burke and Ernzerhof~ \citep{phrL77_3865}, was used.
After self-consistency had been reached in the LDA calculations, the total energy was calculated with Full Charge Density Technique~ \citep{phrB55_13521,levente_book} (FCD).
In this last step the Generalized Gradient Approximation (GGA) was used.
This procedure provides essentially the same materials properties as those of a self-consistent GGA scheme for non-magnetic systems.
While calculations are done for $T=0$ K, the effect of temperature on elastic properties is expected to decrease with increasing pressure~\citep{anderson}.
\\
We have performed the calculations for $Fe$, $Fe_{98}Mg_{02}$, $Fe_{95}Mg_{05}$ and $Fe_{90}Mg_{10}$ in order to investigate the effects of alloying Mg in Fe at small Mg content. 

For the EMTO calculations, the integration in reciprocal space was performed over a grid of 29x29x29 points, and the Green's function was integrated using 16 points in the complex energy plane. 
The basis set used \emph{s,p,d,f} orbitals.

\section{Results} \label{Results}
\begin{figure}[h!]
\includegraphics[width=0.45\textwidth]{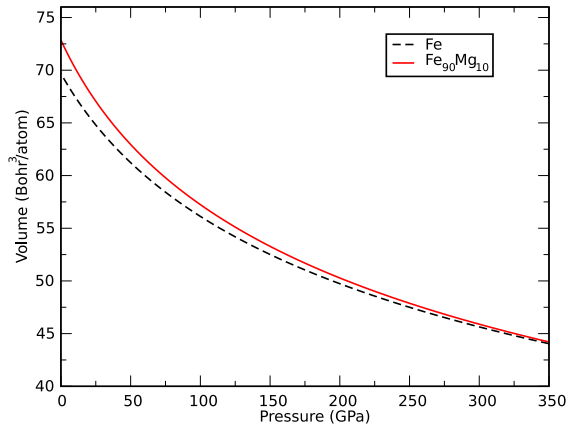}
\caption{\label{fig:V_vs_P}
(Color online) Volume as function of pressure in fcc $Fe_{90}Mg_{10}$ (red full line) and pure Fe (black dashed line).
}
\end{figure}

The volumes as function of pressure for pure Fe and for $Fe_{90}Mg_{10}$ are shown in Fig.~\ref{fig:V_vs_P}.
From this figure it is clear that adding Mg to Fe increases the volume corresponding to a certain pressure. 
This is expected due to the larger atomic size of Mg, and is well-known from earlier work.
When the pressure is increased, the difference in volumes decreases, which is due to the fact that Mg has higher compressibility than Fe. \citep{phrL95_245502} \\

\begin{figure}[h!]
\includegraphics[width=0.45\textwidth]{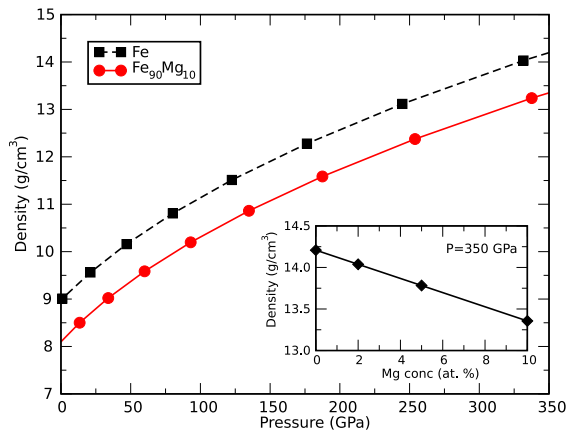}
\caption{\label{fig:density_vs_P}
(Color online) Density as function of pressure in fcc $Fe_{90}Mg_{10}$ (red full line) and pure Fe (black dashed line).
The inset shows the density as function of Mg content for $P=350\, GPa$.
}
\end{figure}

In Fig.~\ref{fig:density_vs_P} we report the density as function of pressure.
The results show that adding $Mg$ into Fe decreases the density, which is expected. 
The results are similar to that reported earlier for hcp Fe-Mg alloys~ \citep{EPSL271_221}.
The density for $Fe_{90}Mg_{10}$ lies close to those of the Preliminary Reference Earth Model (PREM)~ \citep{pepi25_297}.
However, including temperature effects in our calculations is likely to change the agreement.
From the inset, showing the concentration dependence at $P=350$ GPa, we see that adding $2$ \% Mg causes the density to decrease about $1.5$ \%.

\begin{figure}[h!]
\includegraphics[width=0.45\textwidth]{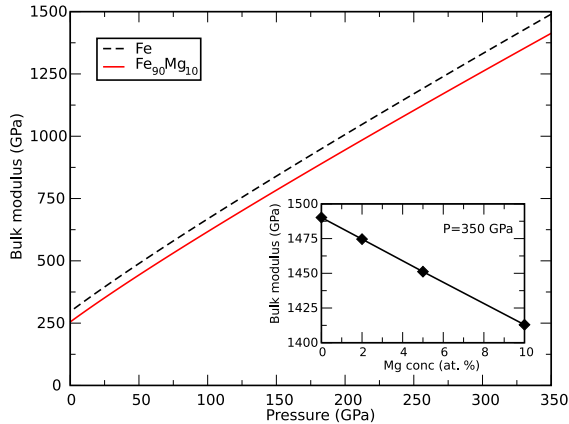}
\caption{\label{fig:B_vs_P}
(Color online) Bulk modulus as function of pressure in fcc $Fe_{90}Mg_{10}$ (red full line) and pure Fe (black dashed line).
The inset shows the bulk modulus as function of Mg content for $P=350\, GPa$.
}
\end{figure}

Next, Fig.~\ref{fig:B_vs_P} shows the bulk modulus as function of pressure.
Also here the results are similar to those reported for hcp~ \citep{EPSL271_221}.
The bulk modulus is about 55 GPa lower for $Fe_{90}Mg_{10}$ than for pure Fe over the whole pressure range considered here, indicating that the FeMg alloy is indeed softer than pure Fe.
Also here we note that the bulk modulus for $Fe_{90}Mg_{10}$ is close to that of the PREM~ \citep{pepi25_297}, but temperature effects may change this, and other elastic properties may not agree to such a large degree.
Still, this work aims to show the effects of adding Mg in Fe, which is still likely to be important at elevated temperatures.
At 350 GPa, adding $2$ \% Mg decreases the bulk modulus by about $1.0$ \%. \\

\begin{figure}[h!]
\begin{center}
\includegraphics[width=0.4\textwidth]{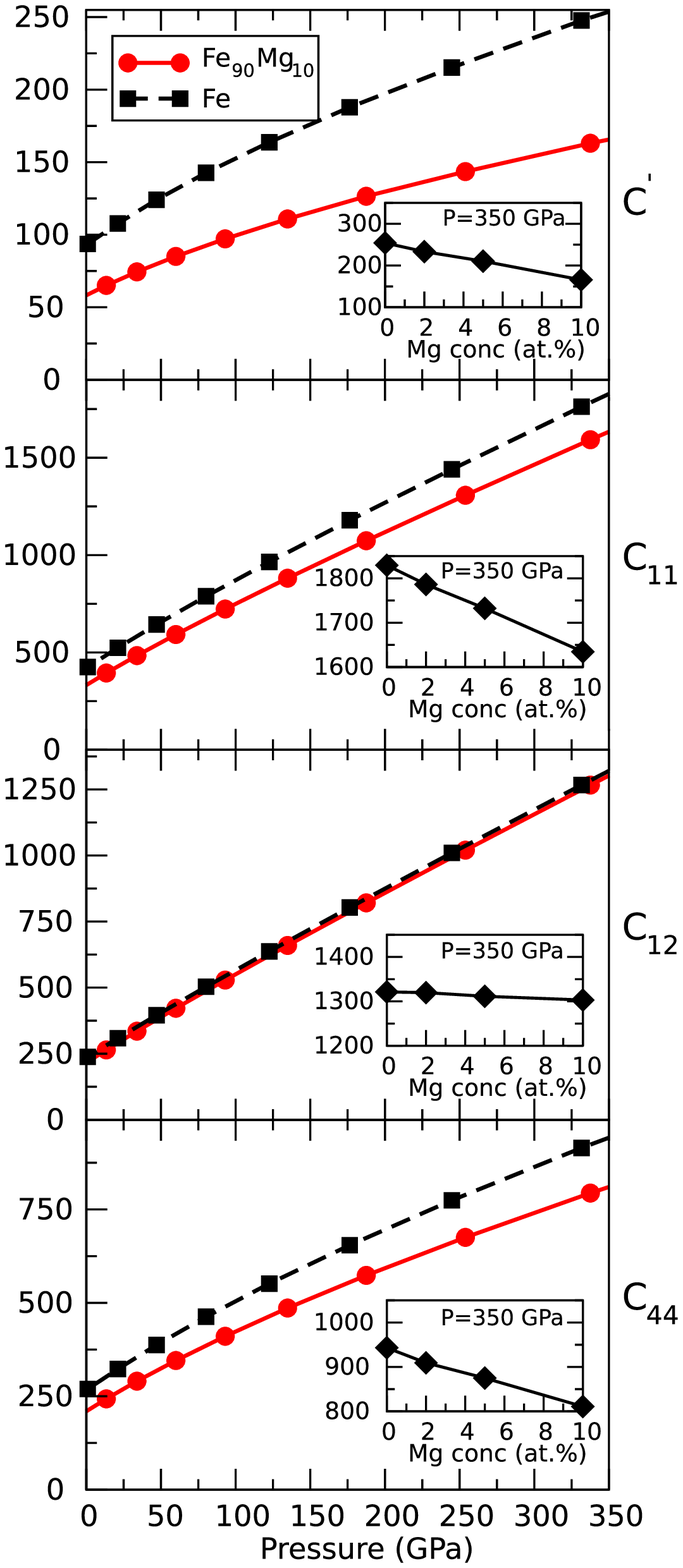}
\caption{\label{fig:elastic-vs-P}
(Color online) Elastic constants as function of pressure in fcc $Fe_{90}Mg_{10}$ (red full line) and pure Fe (black dashed line). All elastic constants are in GPa.
The inset in each subfigure show the corresponding elastic constant as function of Mg content for $P=350\, GPa$. Note that all the insets have the same scale (250 GPa).
}
\end{center}
\end{figure}

The single-crystal elastic constants are shown in Fig.~\ref{fig:elastic-vs-P}.
It is interesting to note that both $c_{11}$ and $c_{44}$ show considerable softening when adding $Mg$.
This is also evident for $c^{\prime}$, which is shown for comparison.
However, the $c_{12}$ elastic constant is almost constant, and for very high pressures it even increases a little when adding $Mg$.
This can be understood from the relations used to obtain $c_{11}$ and $c_{12}$ from $c^{\prime}$ and the bulk modulus:
\begin{eqnarray} \label{eq:c11-c12}
c_{11} = B + \frac{4}{3}c^{\prime} \nonumber \\
c_{12} = B - \frac{2}{3}c^{\prime}
\end{eqnarray}
Since both the bulk modulus and $c^{\prime}$ show similar behavior when adding $Mg$, $c_{11}$ will follow the same trend.
However, Eq.~(\ref{eq:c11-c12}) suggests that for $c_{12}$ the effect is to some extent canceled out.
This can also be seen from the inset for $c_{12}$ in Fig.~\ref{fig:elastic-vs-P}.
For $P=350$ GPa, adding $2$ \% Mg decreases $c^{\prime}$ by about $8.2$ \%, $c_{11}$ by about $2.4$ \%, $c_{12}$ only $0.1$ \% and $c_{44}$ by $3.6$ \%.\\

\begin{figure}[h!]
\includegraphics[width=0.45\textwidth]{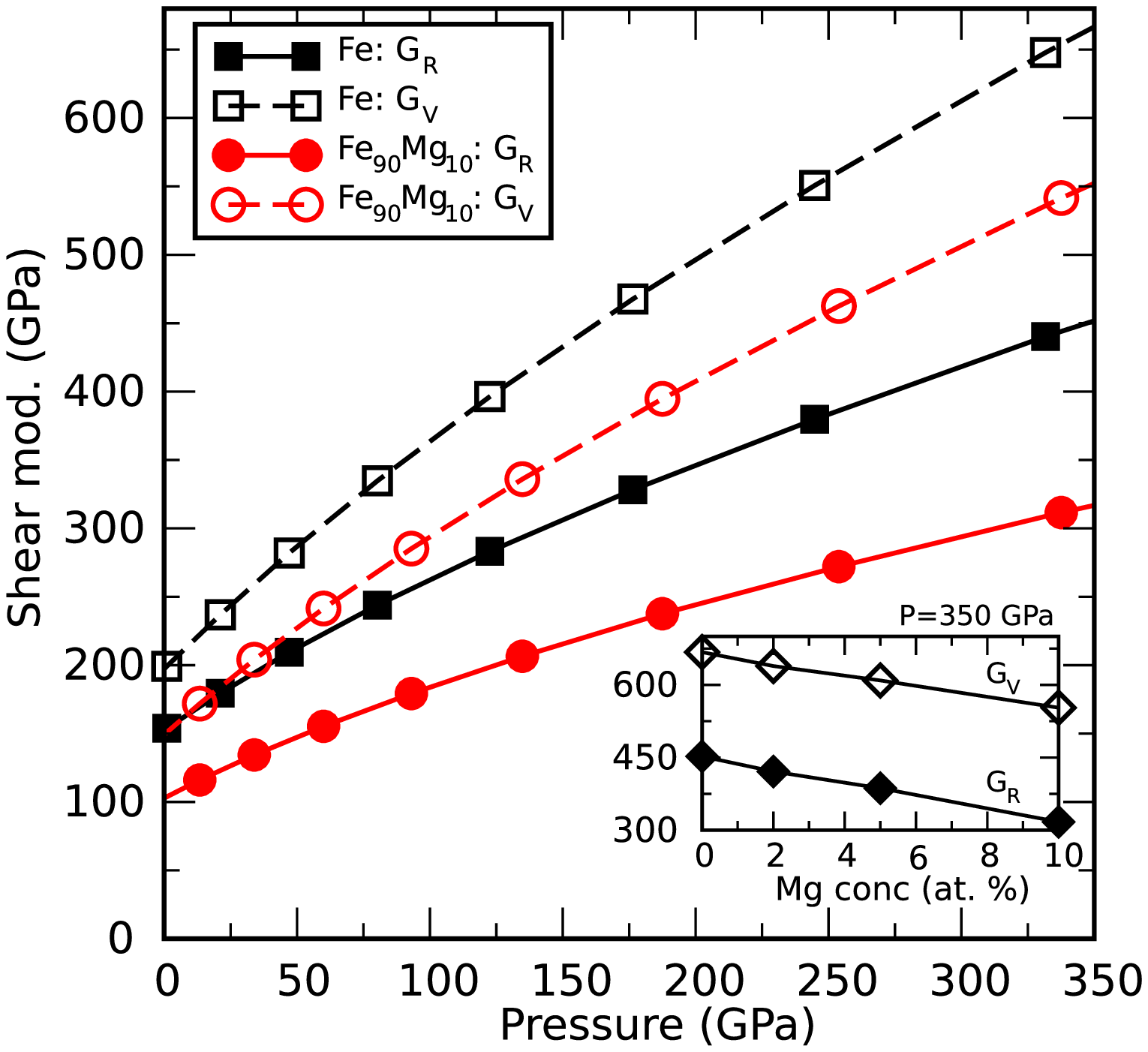}
\caption{\label{fig:GvGr-vs-P}
(Color online) Polycrystalline shear modulus as function of pressure. Black lines and squares denote results for fcc Fe, while red lines and circles denote fcc $Fe_{90}Mg_{10}$. Further, full lines and filled symbols denote $G_R$, while dashed lines and empty symbols denote $G_V$.
The inset shows $G_V$ (open diamonds) and $G_R$ (filled diamonds) as function of Mg content for $P=350\, GPa$.
}
\end{figure}

The polycrystalline elastic constants are shown in Fig.~\ref{fig:GvGr-vs-P}. 
This figure shows that not just the single-crystal elastic constants change considerably when going from pure Fe to $Fe_{90}Mg_{10}$.
A large degree of softening takes place for both $G_R$ and $G_V$ when adding $Mg$ to Fe.
For $P=350$ GPa, adding $2$ \% Mg decrease $G_V$ by $4.3$ \% and $G_R$ by $6.9$ \%.\\

\begin{figure}[h!]
\includegraphics[width=0.45\textwidth]{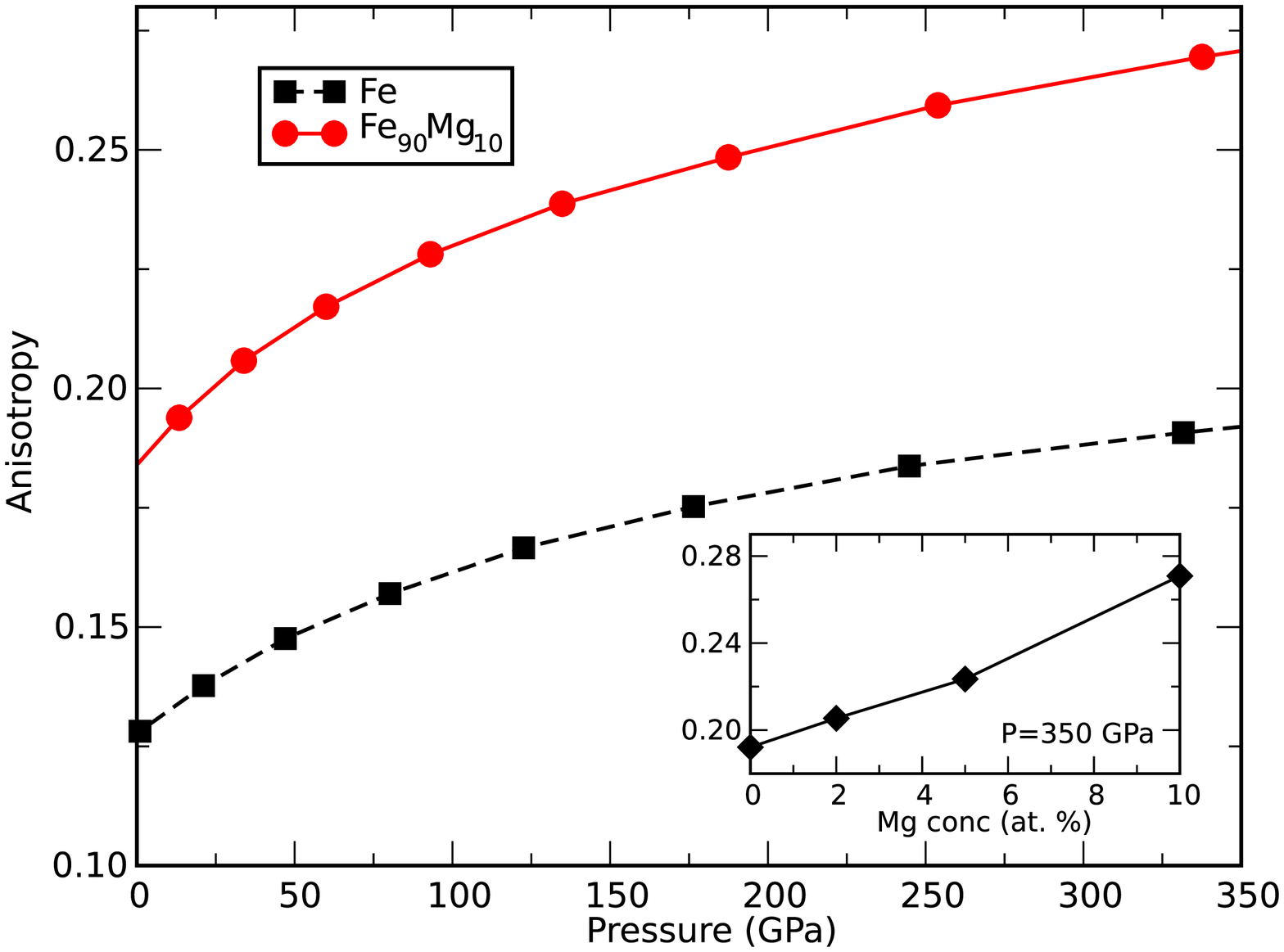}
\caption{\label{fig:anisotropy_Vs_P}
(Color online) Voigt-Reuss-Hill anisotropy as function of pressure in fcc $Fe_{90}Mg_{10}$ (red full line) and pure Fe (black dashed line).
The inset shows the anisotropy as function of Mg content for $P=350\, GPa$.
}
\end{figure}

Fig.~\ref{fig:anisotropy_Vs_P} shows the Voigt-Reuss-Hill elastic anisotropy ($A_{VRH}$) as function of pressure for pure fcc Fe and for fcc $Fe_{90}Mg_{10}$.
This graph shows that adding $10\%$ $Mg$ into Fe cause a remarkable increase of the anisotropy.
At 300 GPa, the anisotropy is about $40 \%$ higher for fcc $Fe_{90}Mg_{10}$ than for pure fcc Fe.
The pressure dependence of the anisotropy is also higher for $Fe_{90}Mg_{10}$ than for pure Fe.

\section{Conclusion} \label{Conclusion}
We have calculated the elastic properties of fcc Fe and $Fe_{90}Mg_{10}$ random alloy up to pressures of the Earth's core from first-principles methods.
We show that adding $10$ at. $\%$ $Mg$ into iron has a substantial effect on the single crystal elastic constants. 
Moreover, even as small Mg concentrations as 2 \% substantially affect some elastic constants, ie $c^{\prime}$ which decreases by 8.2 \% at 350 GPa.
Further we find that the effect on the elastic anisotropy is remarkably high and is higher for the fcc FeMg alloy than for pure fcc Fe.
We therefore conclude that the possibility of Mg in the Earth's core must be included in models, even though it may not be the dominant light element.

\section{Acknowledgments} \label{Acknowledgements}
Useful discussions with L. Vitos are gratefully acknowledged.
We are grateful for financial support from the Swedish research council and the G{\"o}ran Gustafsson Foundation for Research in Natural Sciences and Medicine.
The National Supercomputer centre (NSC) and Center for Parallel Computers (PDC) are acknowledged for computer support.


\end{document}